\documentclass[twocolumn]{revtex4-1}

\usepackage{amssymb,amsthm}
\usepackage[english]{babel}
\usepackage{graphicx}
\usepackage{color}
\usepackage{multirow}
\usepackage{tikz}
\usepackage{float}
\let\oldhat\hat
\renewcommand{\vec}[1]{\mathbf{#1}}
\renewcommand{\hat}[1]{\oldhat{\mathbf{#1}}}

\begin{document}

\title{Simulating flexible polymers in a potential of randomly distributed hard disks}

\author{Sebastian Sch\"obl}
\author{Johannes Zierenberg}
\author{Wolfhard Janke}
\affiliation{Institut f\"ur Theoretische Physik, \\ 
Universit\"at Leipzig, Postfach 100920, D--04009 Leipzig, Germany}

\begin{abstract} 
  We perform equilibrium computer simulations of a two-dimensional pinned
  flexible polymer exposed to a quenched disorder potential consisting of hard
  disks. We are especially interested in the high-density regime of the
  disorder, where subtle structures such as cavities and channels play a central
  role. We apply an off-lattice growth algorithm proposed by Garel and Orland
  [J. Phys. A \textbf{23}, L621 (1990)], where a distribution of polymers is
  constructed in parallel by growing each of them monomer by monomer.  In
  addition we use a multicanonical Monte Carlo method in order to cross-check
  the results of the growth algorithm. We measure the end-to-end distribution
  and the tangent-tangent correlations. We also investigate the scaling behavior
  of the mean square end-to-end distance in dependence of the monomer number.
  While the influence of the potential in the low-density case is merely
  marginal, it dominates the configurational properties of the polymer for high
  densities.
\end{abstract}

\maketitle

\section{Introduction}

Transport phenomena and polymers in porous media
\citep{Guillot,Cannell,PhysRevLett.57.1741} already required theoretical models of
polymers in disordered systems in the 1980s. Since then these
problems have been widely discussed
\citep{Baumgaertner1987,edwards:2435,CatesBall1988,PhysRevA.40.1720,PhysRevA.40.4675,Goldschmidt,Goldschmidt2003,0305-4470-25-23-016,PhysRevA.41.5345,1751-8121-42-1-015001,ThirumalaiPolInPorMed}.
In this paper we apply two algorithms for simulating the equilibrium properties
of a flexible polymer in a disorder potential of hard disks. We are especially
interested in the high-density regime of the disorder, which gives rise to structures
that are hard to tackle with common methods.  We place the disks of the
potential randomly onto the sites of a square lattice, so that,
e.g., the distance between nearest neighbors can be controlled. This somewhat artificial
arrangement, which causes, e.g., narrow channels and small cavities (see
Sec.~\ref{sec:disorderPotential}), was chosen in order to investigate the
influence of those structures on flexible polymers and to test
the methods we use. In forthcoming work we want to apply them to the more general biophysically inspired problem
of semiflexible polymers in crowded media where the disorder is irregular and
possibly correlated (e.g., a hard-disk fluid) \cite{SSchoebl}. We access this
problem by applying and comparing two algorithms. One algorithm is an
off-lattice growth algorithm \cite{Orland1990}. Growth algorithms are
intensively used for lattice polymer systems
\cite{PhysRevE.56.3682,PhysRevLett.95.058102,0295-5075-82-6-66006,PhysRevLett.101.125701} whereas they
are rarely applied for the off-lattice case. The other method is the
multicanonical Monte Carlo method \cite{Berg-muca1,
Berg-muca2,MulticanonicalMCsimulations}, which is a common instrument for
handling systems with rough energy landscapes. It has already proven to be very
efficient for polymeric systems \citep{Junghans2006}.  Throughout our analysis
we found perfect agreement of the two methods except for some special case of
parameters. We will discuss the merits and drawbacks of these methods in Sec.~\ref{sec:resAndDisc}. 

The rest of the paper is organized as follows: In Sec.~\ref{sec:polymerModelAndDis}
we describe our polymer model and how the disorder is realized. Then, in
Sec.~\ref{sec:algorithms}, we describe the algorithms. At the end of
this section, we specify the parameters for our simulations. In
Sec.~\ref{sec:resAndDisc} we show our findings. A conclusion is given in Sec.~\ref{sec:conc}.

\section{Polymer model and disorder}\label{sec:polymerModelAndDis}

We simulate a polymer exposed to a disorder landscape consisting of hard disks. The
polymer is pinned at one end. We carry out the quenched disorder average as
follows: We choose a random starting point for each disorder realization and run
an equilibrium computer simulation. We estimate averages from the
resulting distribution of polymer configurations. Averages for a single
disorder realization are written in angular brackets $\langle ... \rangle$. This is done for all disorder
realizations and the quenched average is calculated from this by averaging
over the measured values of the single disorder realizations. The quenched
average is written as $[\langle ... \rangle]$. The parameters we
use for our simulations are described in detail in Sec.~\ref{sec:simulationParameters}. 

\subsection{Polymer model}

Our polymer model is a freely jointed chain. Effectively, this is a
bead-stick model whose contour is defined by $N+1$ beads at positions $\vec r_i$
connected by bonds of fixed length $b$.  Therefore the contour has the fixed
length $L=Nb$. The polymer chain is a phantom chain, which means that there
is no monomer-monomer interaction except for the fixed distance between bonded
monomers.  The connecting line of bonded monomers defines unit vectors $\vec t_i
= (\vec r_{i+1}-\vec r_i)/b$. Our methods can easily be adapted to other polymer
models. In a forthcoming work, we will extend it to involve bending energy.

\subsection{Disorder potential}\label{sec:disorderPotential}

The background potential consists of hard disks with diameter $\sigma_i$,
which interact with the monomers of the polymer via hard-core repulsion.  The
interaction potential between monomers and disks is thus described by 
\begin{equation}
  V = \left\{
  \begin{array}{l l}
    \infty & \quad \text{for} \quad d < {\sigma_i}/{2} \\ 0 & \quad \text{else}
  \end{array}\right. 
\end{equation}
where $d$ is the distance between a monomer and the center of a disk
of the background potential (no monomer volume). 

The disks are placed onto the sites of a square lattice with a lattice
constant chosen to be of the order of the disk diameter. This arrangement was
chosen in order to be able to control the distance between neighboring disks. In
such a disorder landscape the algorithms can be well tested. The algorithms
can then easily be applied to other potentials such as, e.g., hard-disk fluids. 

\begin{figure}[t]
  \includegraphics[scale=0.4]{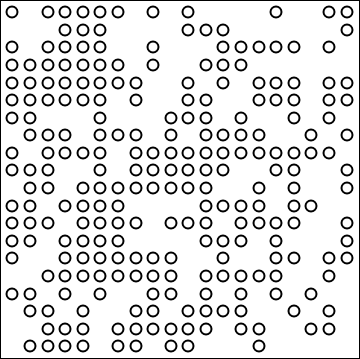}
  \caption{Hard-disk disorder configuration with site occupation probability $p=0.64$.}
  \label{fig:potentialLatticeFluid}
\end{figure}

In order to generate random configurations we occupy each lattice site
with the same probability. The resulting structure mimics -- to a certain extent -- the
structure of a diluted square lattice. Figure~\ref{fig:potentialLatticeFluid} shows an
example configuration of hard disks with site occupation probability $p=0.64$.

\section{Algorithms}\label{sec:algorithms}

\begin{figure*}[!htbp]
  \begin{tabular}{ccccc}
    \begin{minipage}{0.2\textwidth}
      \includegraphics{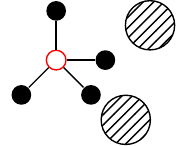}
    \end{minipage}
    &
    \begin{minipage}{0.2\textwidth}
      \includegraphics{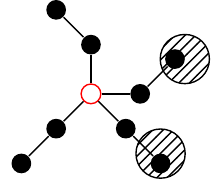}
    \end{minipage}
    &
    \begin{minipage}{0.2\textwidth}
      \includegraphics{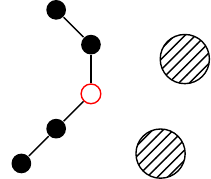}
    \end{minipage}
    &
    \begin{minipage}{0.2\textwidth}
      \includegraphics{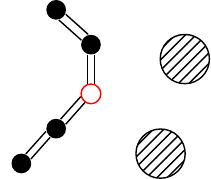}
    \end{minipage}
    &
    \begin{minipage}{0.2\textwidth}
      \includegraphics{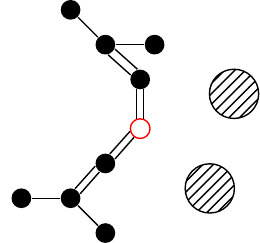}
    \end{minipage}
    \\
    (a)
    &
    (b)
    &
    (c)
    &
    (d)
    &
    (e)
  \end{tabular}
  \begin{minipage}{\textwidth}
    \caption{\label{fig:growthAlg}(Color online) (a) Initial growth in four different directions
    corresponds to four different polymer chains of length 1 starting from the seed
    marked by the red open circle. (b) Each of the chains is
    grown by one monomer. Two of the chains overlap with a disk of the background
    potential. The weight of these chains is 0. (c) The chains with weight 0 are
    removed from the population. (d) The population control parameter keeps the
    overall number of chains approximately constant. On average each of the remaining chains is thus
    replicated and now exists twice in the population. (e) Each of the chains is
    grown independently by one more monomer.}
  \end{minipage} 
\end{figure*}
Application of a standard Markov chain Monte Carlo (MCMC) method based upon the
Metropolis algorithm \citep{Metropolis1953}, that is, building up an initial
polymer chain and then updating it in order to sample the configuration space,
did not work well for our purposes. The updates included only the changes of
angles between neighboring bonds, but even more refined update moves such as
pivot moves would certainly not work efficiently in this situation. For a highly dense
background, these methods result in either a diverging autocorrelation time or
a rejection rate that makes it impossible to create enough configurations to
compute observables.

\subsection{Method 1: Chain growth algorithm}

Following the approach proposed by Garel and Orland \citep{Orland1990}, we
generate ensembles of pinned polymers by placing $M\sim 10^5$ seeds at a
randomly chosen site and simultaneously growing them, monomer by
monomer, until the desired degree of polymerization is reached. After each
growth step, we attain thermal equilibration by replicating or deleting chains
according to their Boltzmann weight (0 if the chain happens to collide with a
background obstacle, 1 otherwise). To avoid an exponential decline in the chain
population, a population control parameter is introduced that keeps the total
number of ensemble members approximately constant (see
Fig.~\ref{fig:growthAlg}). Prior to the proper growth process, 
the overall number of chains in this step is estimated. The population
control parameter is the ratio of the initial number of seeds and the estimated
number of chains in the next step. The weight for each chain is multiplied by
the population control parameter. 
Other types of polymer interaction such as bending energy can be treated
similarly by using different Boltzmann weights.

\subsection{Method 2: Multicanonical Monte Carlo algorithm}

In order to avoid trapping problems of the polymer in a system with hard disks,
we can treat the excluded areas as finite potentials. That way, the polymer is allowed 
to access the previously forbidden disks with the Boltzmann probability, depending on
the energy penalty from monomers located on the disks.
This approach allows for a multicanonical simulation \cite{Berg-muca1, Berg-muca2} in
the amplitude parameter of the disk potential, opening the possibility of
reproducing the limiting cases of a free polymer (zero amplitude) and a polymer
in hard disk disorder (infinite amplitude).

In general, the total energy of a polymer may be defined as the sum of the intra-polymer energy
$E_{\rm poly}$ and the potential energy $k E_{\rm pot}$, where $E_{\rm pot}$ is the sum of 
monomers located on disks and $k$ is the amplitude of the disk potential in energy units.
While the intra-polymer energy is a continuous quantity, the potential energy is discrete.
The canonical partition function is given as the sum over all polymer configurations $\{x_i\}$
\begin{equation} 
  Z_{\rm can} = \sum\limits_{\{x_i\}} e^{-\beta\big(E_{\rm poly}(\{x_i\}) + k
  E_{\rm pot}(\{x_i\})\big)}.
  \label{eq:ZCan}
\end{equation}
%
In our case, this simplifies even more as we only consider
a flexible polymer with vanishing intra-polymer energy $E_{\mathrm{poly}}$, but the method also works for polymers 
with interactions.
It is possible to separate the Boltzmann factor into contributions of the 
intra-polymer energy and of the potential energy. Replacement of the Boltzmann factor of the potential
energy by a variable weight factor results in our multicanonical partition function
\begin{equation}
  Z_{\rm muca} = 
  \sum\limits_{\{x_i\}} e^{-\beta E_{\rm poly}(\{x_i\})} W\big(E_{\rm pot}(\{x_i\})\big). 
\end{equation}
Independent on the choice of the weights, the canonical expectation values can always be recovered by
\begin{equation}
  \langle O \rangle_{\rm can} = \frac{ \left\langle O W^{-1}\big(E_{\rm pot}(\{x_i\})\big) e^{-\beta k E_{\rm pot}(\{x_i\})} \right\rangle_{\rm muca} }
  { \left\langle   W^{-1}\big(E_{\rm
  pot}(\{x_i\})\big) e^{-\beta k E_{\rm
  pot}(\{x_i\})} \right\rangle_{\rm muca} }.
\end{equation}

Now, the weights may be adjusted such that states that initially occur
frequently are suppressed, while states with rarely occurring potential energies
are amplified. The weights are iterated in equilibrium simulations until the
resulting histogram of the potential energy $H(E_{\rm pot})$ is flat and thus
configurations with different numbers of monomers located on hard disks appear
with the same rate, allowing the polymer to cross over previous barriers of hard
disks. This may be achieved in different ways
\cite{MulticanonicalMCsimulations}.  In our case we start with the Boltzmann
weights in Eq.~(\ref{eq:ZCan}) with $\beta k=1$ and after each equilibrium
simulation we recalculate the weights with
\begin{equation} W^{(n+1)}(E_{\rm pot}) = \frac{W^{(n)}(E_{\rm pot})}{H^{(n)}(E_{\rm pot})}.
\end{equation}
This simple weight update already leads to a quick convergence to flat histograms. As the first
histogram may be narrow, it is of advantage to begin with small statistics,
increasing the number of updates in each iteration upon a chosen threshold.

In the end, the resulting weights are used to perform a final simulation.
The desired observables are obtained by reweighing the final time series,
meaning that the weights with which the observables were measured
are replaced by the weights with which they would appear in the canonical ensemble.
With proper normalization this gives
\begin{equation} \langle O \rangle_{\mathrm{can}} \approx
  \overline{O}_{\mathrm{can}} = \frac{\sum\limits_i \frac{e^{-\beta k
  E_{{\rm pot},i}}}{W(E_{{\rm pot},i})} O_i }{\sum\limits_i \frac{e^{-\beta k
  E_{{\rm pot},i}}}{W(E_{{\rm pot},i})} }.  
\end{equation}

\section{Simulation parameters}\label{sec:simulationParameters}

\begin{figure}[h]
  \begin{tabular}{ccc}
    \begin{minipage}{0.15\textwidth}
      \includegraphics[scale=0.22]{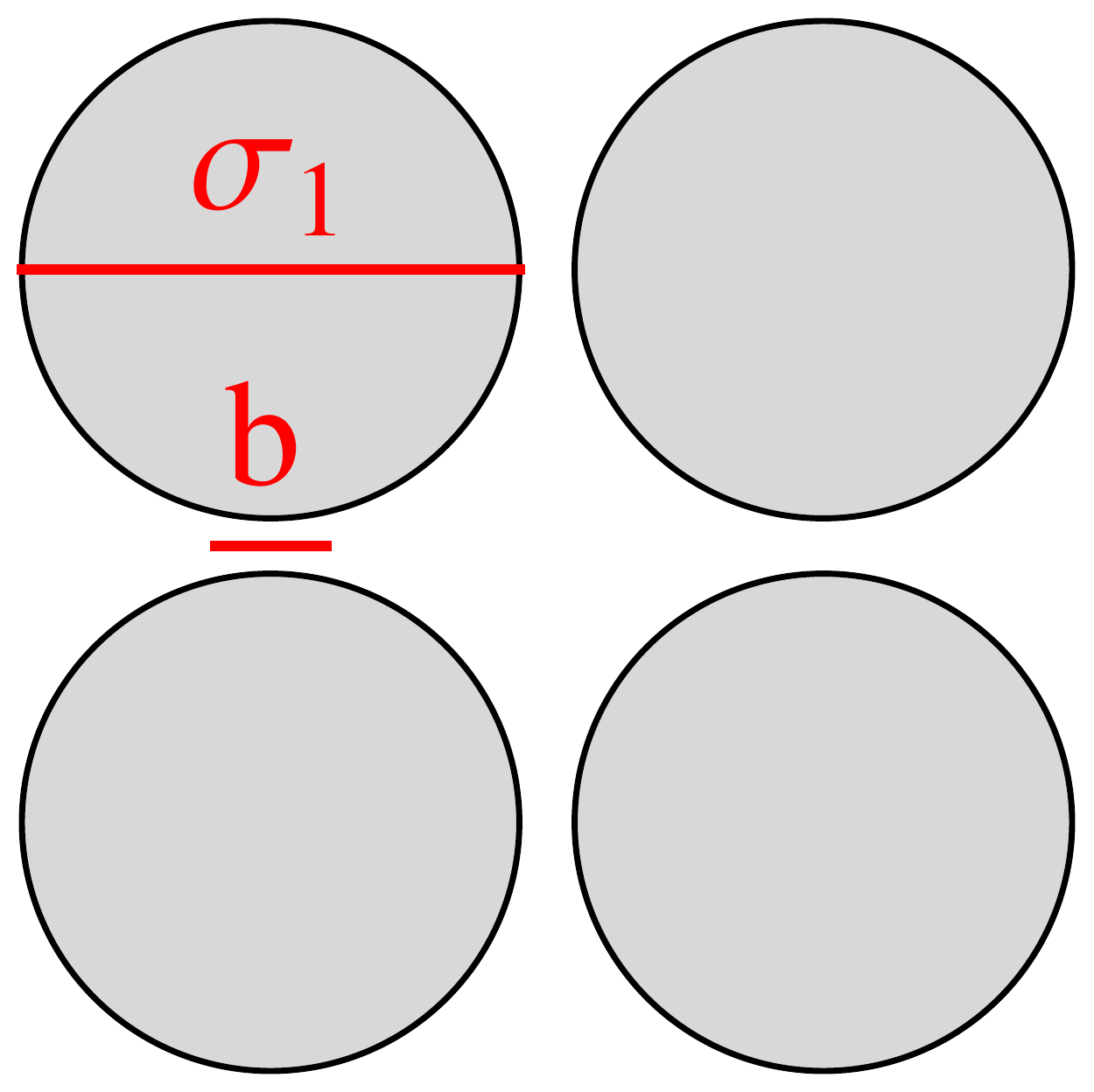}
    \end{minipage}
    &
    \begin{minipage}{0.15\textwidth}
      \includegraphics[scale=0.22]{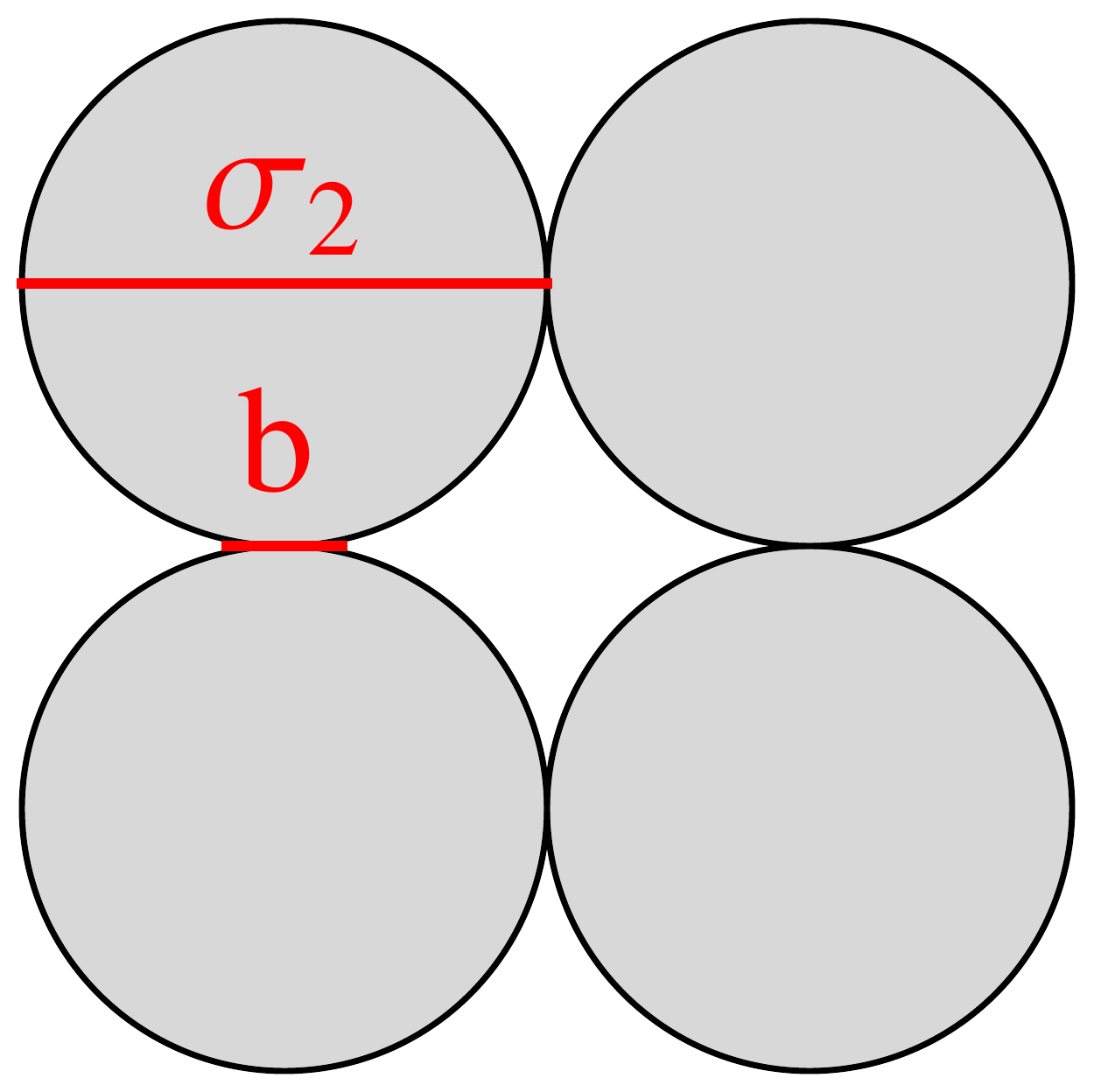}
    \end{minipage}
    &
    \begin{minipage}{0.15\textwidth}
      \includegraphics[scale=0.22]{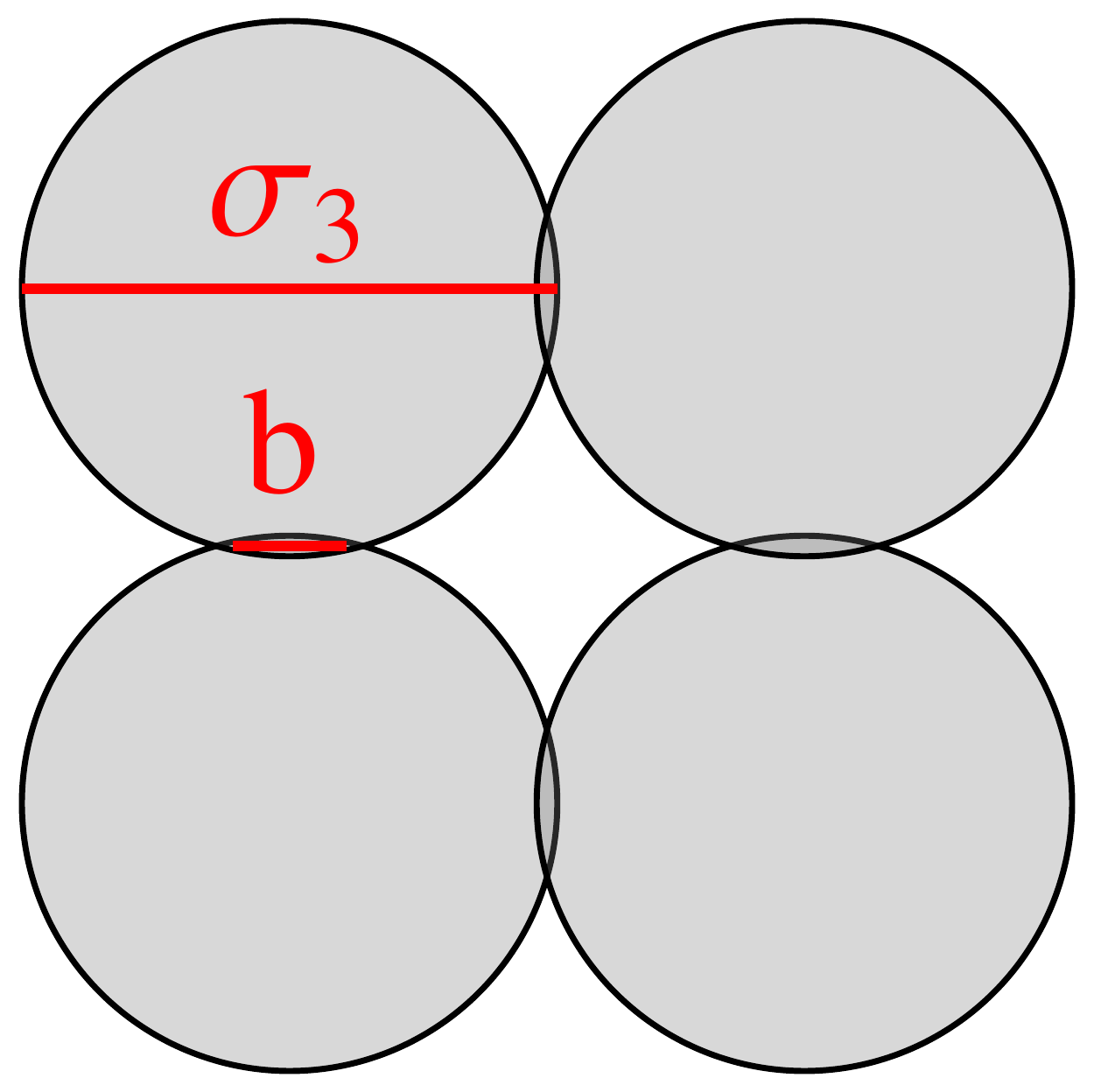}
    \end{minipage} \\
    (a) & (b) & (c)
  \end{tabular}
  \caption{(Color online) Sketch of the different disk sizes $\sigma_i$ for the background potential.
  $b$ is the bond length. 
  For case (a), which belongs to $\sigma_{1}$, there is a finite
  channel between neighboring disks. (b) shows the case where neighboring
  disks touch each other at one point (case $\sigma_{2}$). As there are
  no bond and no monomer volume, there is still a small probability of crossing this
  touching point. For (c) this is no longer possible as the overlap width of
  neighboring disks is larger than the bond length (case $\sigma_{3}$).}
  \label{fig:hardDiskSketch}
\end{figure}
We simulate in a square of fixed area $A=1$ with periodic boundary
conditions. The diameter of the disks is set to $\sigma_{\rm 1}=0.045$,
$\sigma_{\rm 2}=0.05$ and $\sigma_{\rm 3}=0.051$. The disks are placed onto the
sites of a square lattice with $20\times 20$ sites and lattice constant
$a=0.05$.  The site occupation probabilities of the lattice include $p=0, 0.13,
0.25, 0.38, 0.51, 0.64, 0.76, 0.89, 1.00$. For $\sigma=a$, these densities
correspond on average to the area fractions $\rho= 0, 0.1, 0.2, 0.3, 0.4, 0.5, 0.6, 0.7, 0.785$.
The number of monomers---except for some scaling considerations---per polymer
chain is $N+1=30$. The monomers are considered pointlike. 
The bond length is set to $0.01$. Accordingly, the only constraints are the fixed
length of the bonds and the fact that a monomer is not allowed to be placed on a disk of
the background potential.  If we look more closely, we realize that for
$\sigma_{\rm 1}$ there is a channel of half the bond length between neighboring
disks [see Fig.~\ref{fig:hardDiskSketch}(a)]. For $\sigma_{\rm 2}$
[Fig.~\ref{fig:hardDiskSketch}(b)], neighboring disks of the background
potential touch each other at one point. The bonds of the polymer can overlap
with the disks of the background potential. As there are no monomer and no bond
volume, there is a small probability of the polymer getting through the touching
point of two neighboring disks.  For $\sigma_{\rm 3}$
[Fig.~\ref{fig:hardDiskSketch}(c)] this is no longer possible as the overlap
width of neighboring disks is larger than the bond length of the polymer. This
was chosen in order to compare the algorithm's ability to explore narrow
channels in a high-density disorder landscape. For the case of thirty monomers,
the polymer has a length of about six disk diameters if it is completely
stretched. 

\section{Results and discussion}\label{sec:resAndDisc}

\begin{figure*}[!ht]
  \begin{tabular}{cc} 
    \multicolumn{2}{c}{
    \begin{minipage}{0.5\textwidth}
      \includegraphics[scale=0.4]{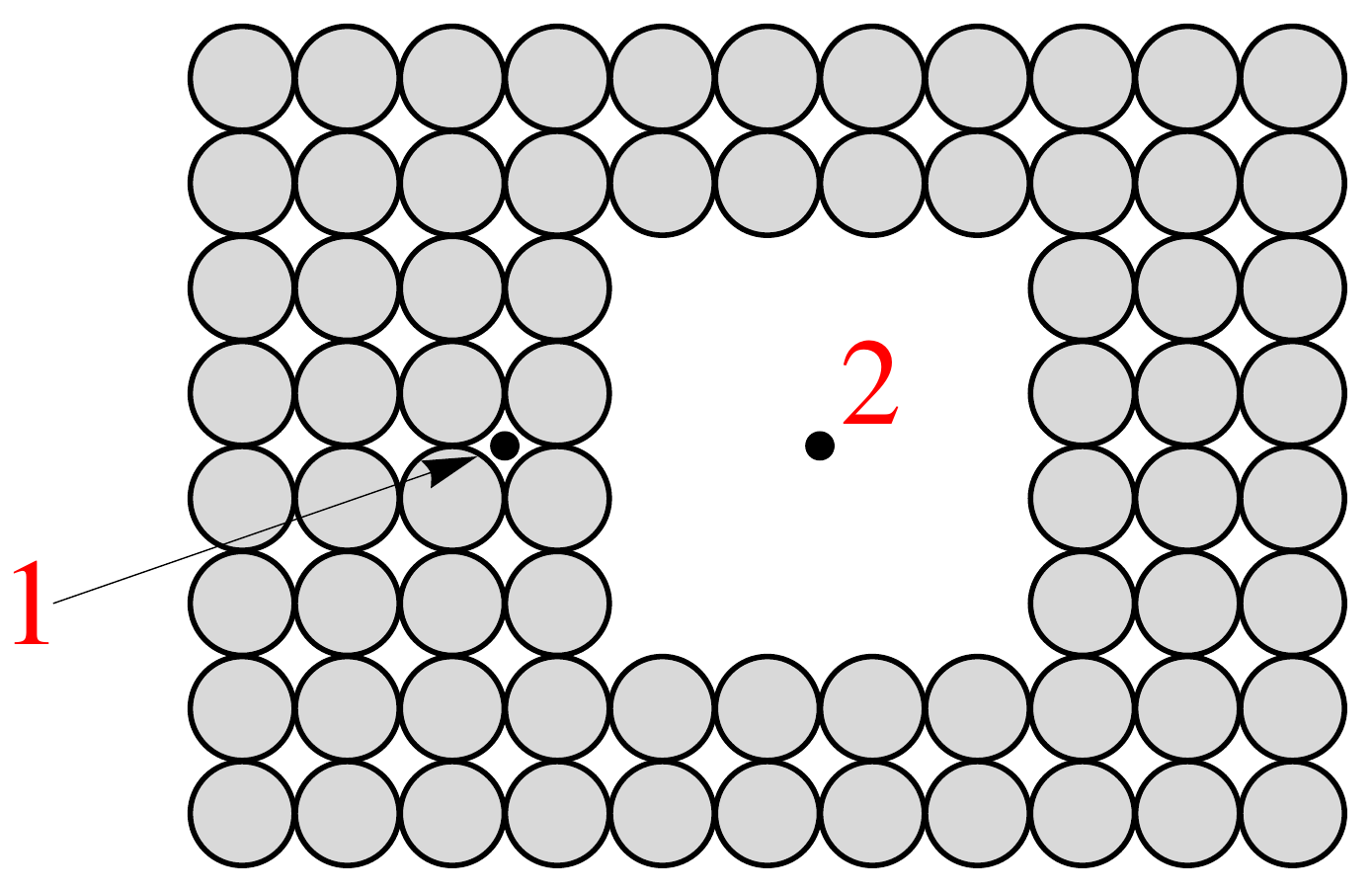}
    \end{minipage} } \\ \multicolumn{2}{c}{ (a)} \\ \begin{minipage}{0.5\textwidth}
      \includegraphics[scale=1]{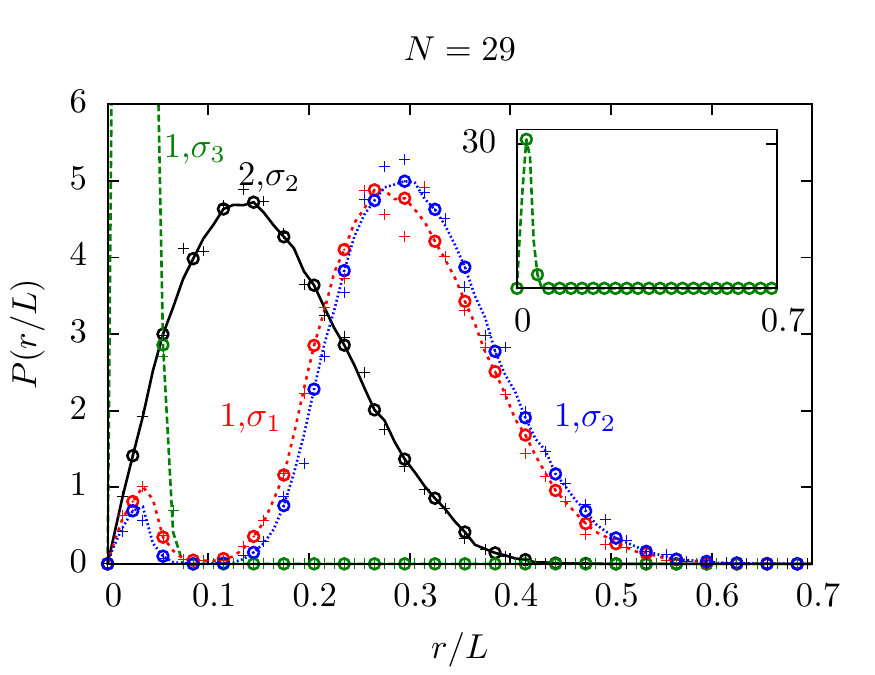}
    \end{minipage} & \begin{minipage}{0.5\textwidth}
      \includegraphics[scale=1]{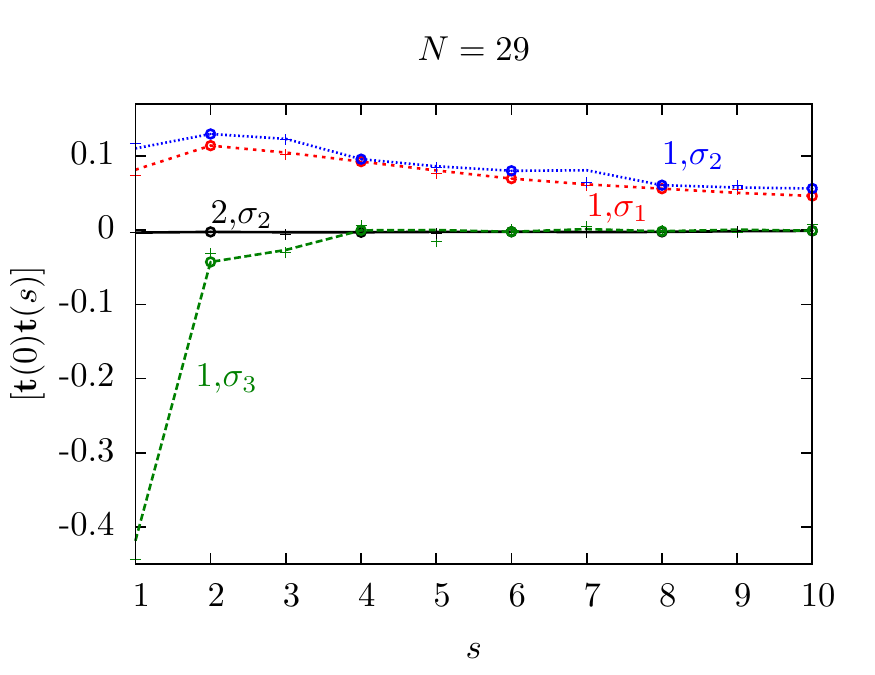}
    \end{minipage} \\ (b) & (c) 
  \end{tabular} 
  \begin{minipage}{\textwidth}
    \caption{\label{fig:singleDisorderConfigurationAnalysis}(Color online) (a) shows a
    distribution of disks with two exemplary pinpoints for polymers. All sides are
    continued with disks. The boundary conditions are periodic. (b) is the accordant
    end-to-end distribution and (c) the
    tangent-tangent correlation function (single simulation; no disorder average).
    $\circ$ shows the data from the growth algorithm, $+$ from the multicanonical
    simulation. The double-peaked curves [dotted (blue) and short-dashed (red)] in (b) as well as the
    single peaked curve [dashed (green)], which is shown in full $y$ range in the inset, belong
    to pinpoint~1. The double peaked curves belong to $\sigma_1$ [short-dashed
    (red)] and $\sigma_2$ [dotted (blue)]. The strongly peaked curve [dashed
    (green)] belongs to $\sigma_3$. The solid (black) curve
    with the broad single peak belongs to pinpoint~2 for $\sigma_2$. The curves for $\sigma_1$ and
    $\sigma_3$ are not shown for pinpoint~2 as they behave similarly to $\sigma_2$.
    The line coding is the same for the tangent-tangent correlations (c).}
  \end{minipage} 
\end{figure*}
In our analysis we focus on three observables: the end-to-end
distribution $P(r)$, the tangent-tangent correlations $\langle\vec{t}(0) \vec{t}(s)\rangle$
and the mean squared end-to-end distance $\langle R_{\mathrm{ee}}^2 \rangle(N)$.
The end-to-end distribution gives the probability for finding a certain
end-to-end distance $r$. 
For a free flexible polymer, the end-to-end distribution is of the form
$P(\widehat{r})\propto \widehat{r} e^{-\widehat{r}^2/2\sigma^2}$, where
$\widehat{r}=r/L$ with $L=b N$.
The tangent-tangent correlation function shows the average correlation of two
bonds $\vec{t}(i)$ and $\vec{t}(i+s)$. In our case it is defined by 
\begin{equation}
  \langle \vec{t}(0)\vec{t}(s) \rangle = \sum_{i=0}^{N-1-s}{\frac{\vec{t}(i)\vec{t}(i+s)}{N-1-s}}.
\end{equation}
The tangent-tangent correlation function is a measure of the stiffness of the
polymer. For a completely flexible polymer, the correlations are zero. The
surrounding disorder can lead to both correlations and anti-correlations
as can be seen in Fig.~\ref{fig:singleDisorderConfigurationAnalysis}(c).

The last observable that we consider is the mean square end-to-end distance in
dependence on the polymer length counted in numbers of bonds. In order to compare
to the literature, we consider the mean square end-to-end distance without normalization.
For free polymers it grows linearly in $N$. 
\begin{figure}\tt
  \includegraphics[scale=0.5]{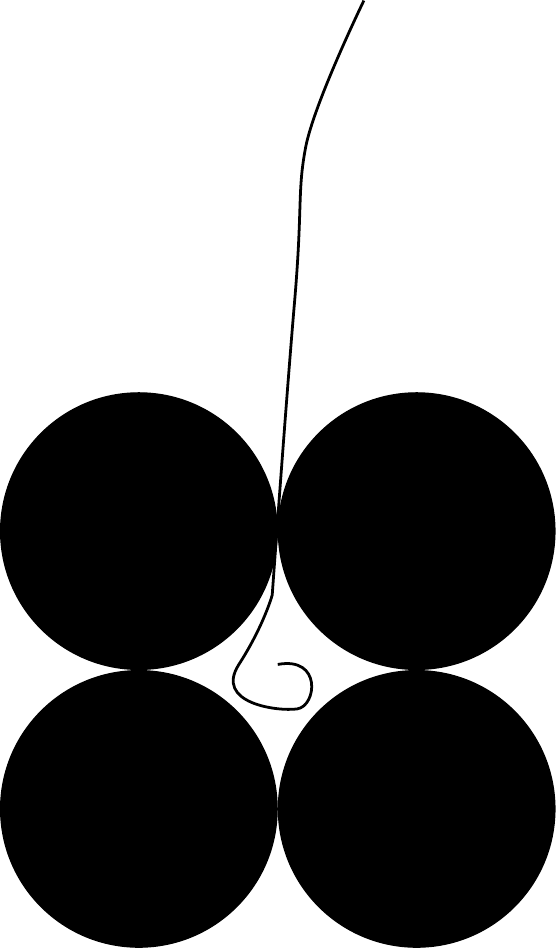} \caption{The sketch of a polymer that finds its way through a narrow
  channel to explore the large space behind.} \label{fig:polymerDiffusesOut}
\end{figure}

The statistical errors are estimated in the standard way by calculating the variance over the disorder
realizations, which are uncorrelated. In the plots we omitted the error bars
as they turned out to be smaller than the plot markers.

\begin{figure*}\tt
  \begin{tabular}{ccc}
    \begin{minipage}{0.33\textwidth}
      \includegraphics[scale=1]{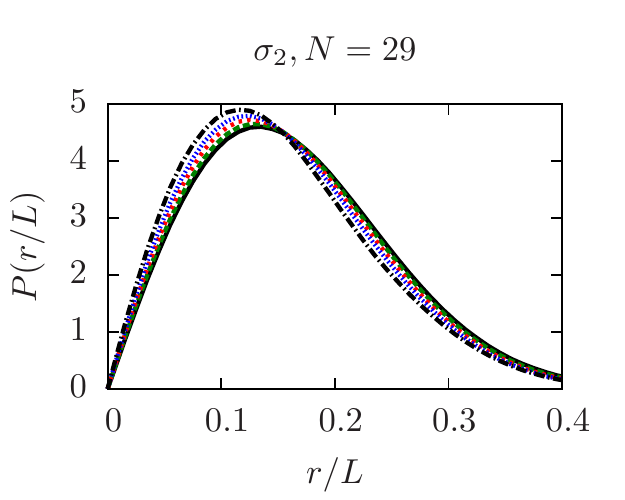}
    \end{minipage} & 
    \begin{minipage}{0.33\textwidth}
      \includegraphics[scale=1]{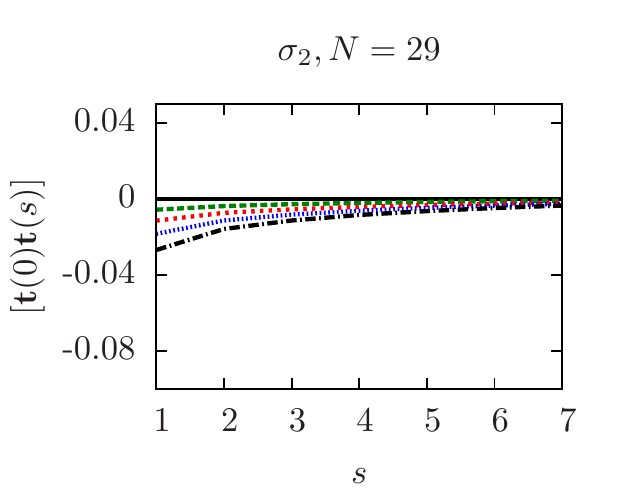}
    \end{minipage} &
    \begin{minipage}{0.33\textwidth}
      \includegraphics[scale=1]{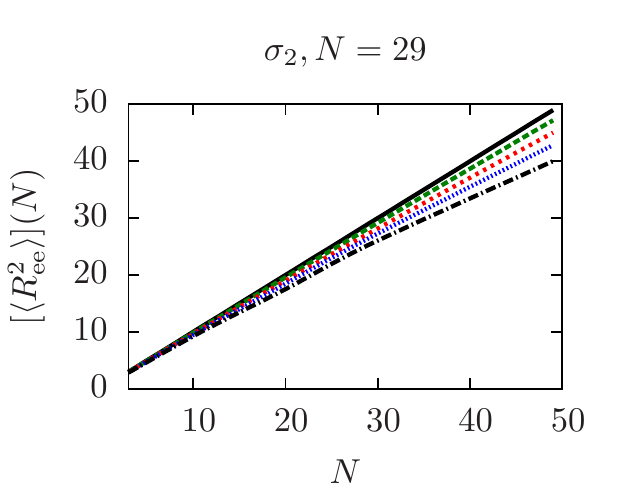}
    \end{minipage}\\
    (a) & (b) & (c)
  \end{tabular}
  \begin{minipage}{\textwidth}
    \caption{\label{fig:endToEndGrowthMucaLowDens}(Color online) (a) End-to-end distribution
    function for site occupation $p=0$
    (\hspace{-0.2cm}\protect\includegraphics[scale=1]{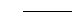}, black
    solid), $0.13$
    (\hspace{-0.2cm}\protect\includegraphics[scale=1]{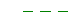},
    green), $0.25$
    (\hspace{-0.2cm}\protect\includegraphics[scale=1]{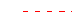}, red), $0.38$
    (\hspace{-0.2cm}\protect\includegraphics[scale=1]{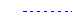}, blue),
    and
    $0.51$
    (\hspace{-0.2cm}\protect\includegraphics[scale=1]{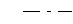},
    black) for increasing peak height. The curves are interpolating lines through the data (whose markers
    have been omitted for better visibility). The results of the two algorithms
    agree within the line thickness. The influence of
    the disk diameter $\sigma_i$ is negligible in this density regime and chosen
    here to be $\sigma_2$. (b) and (c) are the corresponding plots for the
    tangent-tangent correlations and the mean square end-to-end distance (in units
    of squared bond length $b^2$).} 
  \end{minipage}
\end{figure*}
Figure~\ref{fig:singleDisorderConfigurationAnalysis} is a showcase of different
scenarios that can occur during the disorder averaging. Pinpoint~1 is in a small
cavity that is entropically unfavorable for the polymer compared to a larger
space such as can be seen around pinpoint~2. As long as the polymer has the chance
to explore a larger area by escaping from a small cavity through a channel, this
will happen even if the channel is extremely narrow
(Fig.~\ref{fig:polymerDiffusesOut}). 
\begin{figure}\tt
  \begin{tabular}{ccc}
    \begin{minipage}{0.15\textwidth}
      \includegraphics[scale=0.15]{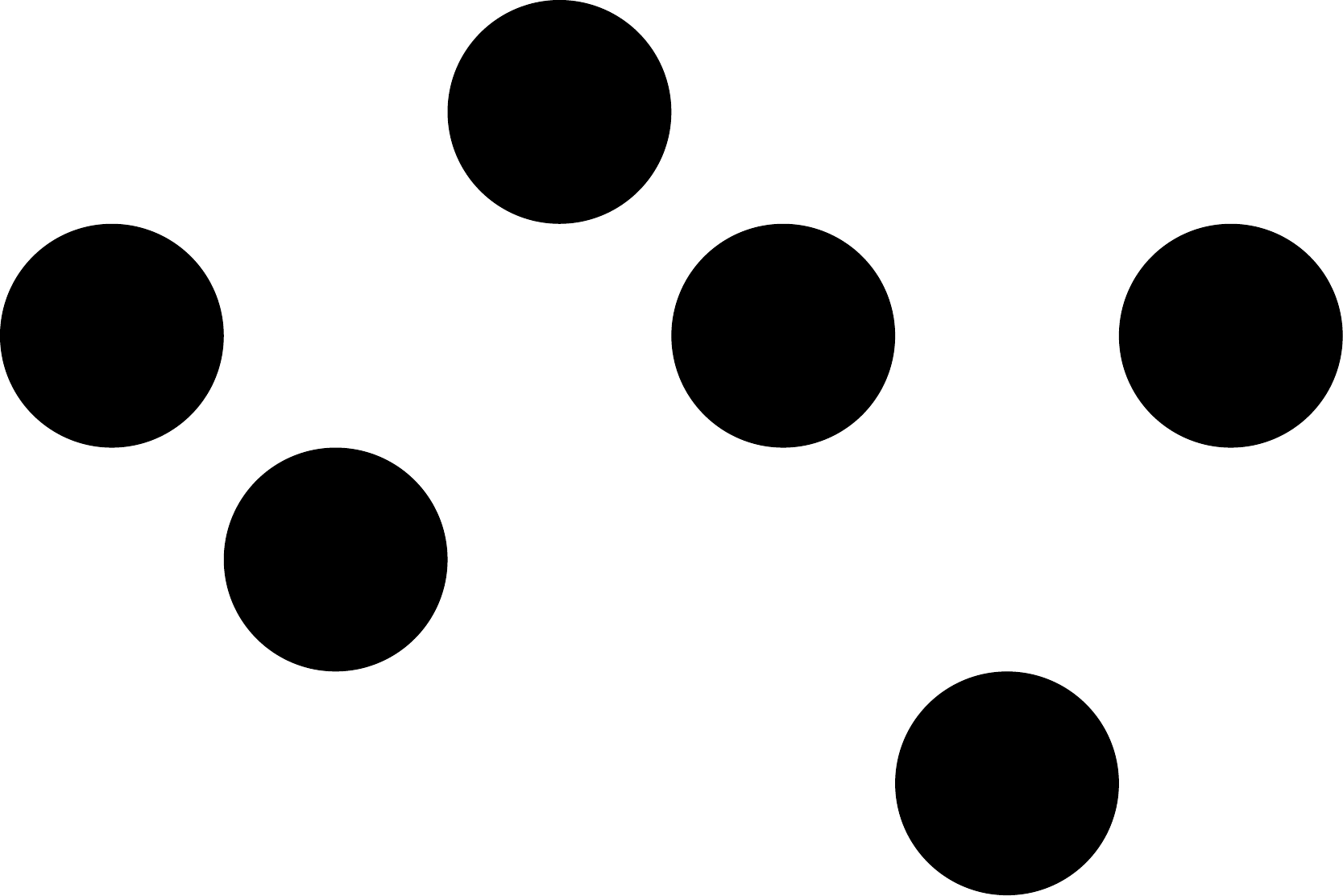}
    \end{minipage} & 
    \begin{minipage}{0.15\textwidth}
      \includegraphics[scale=0.15]{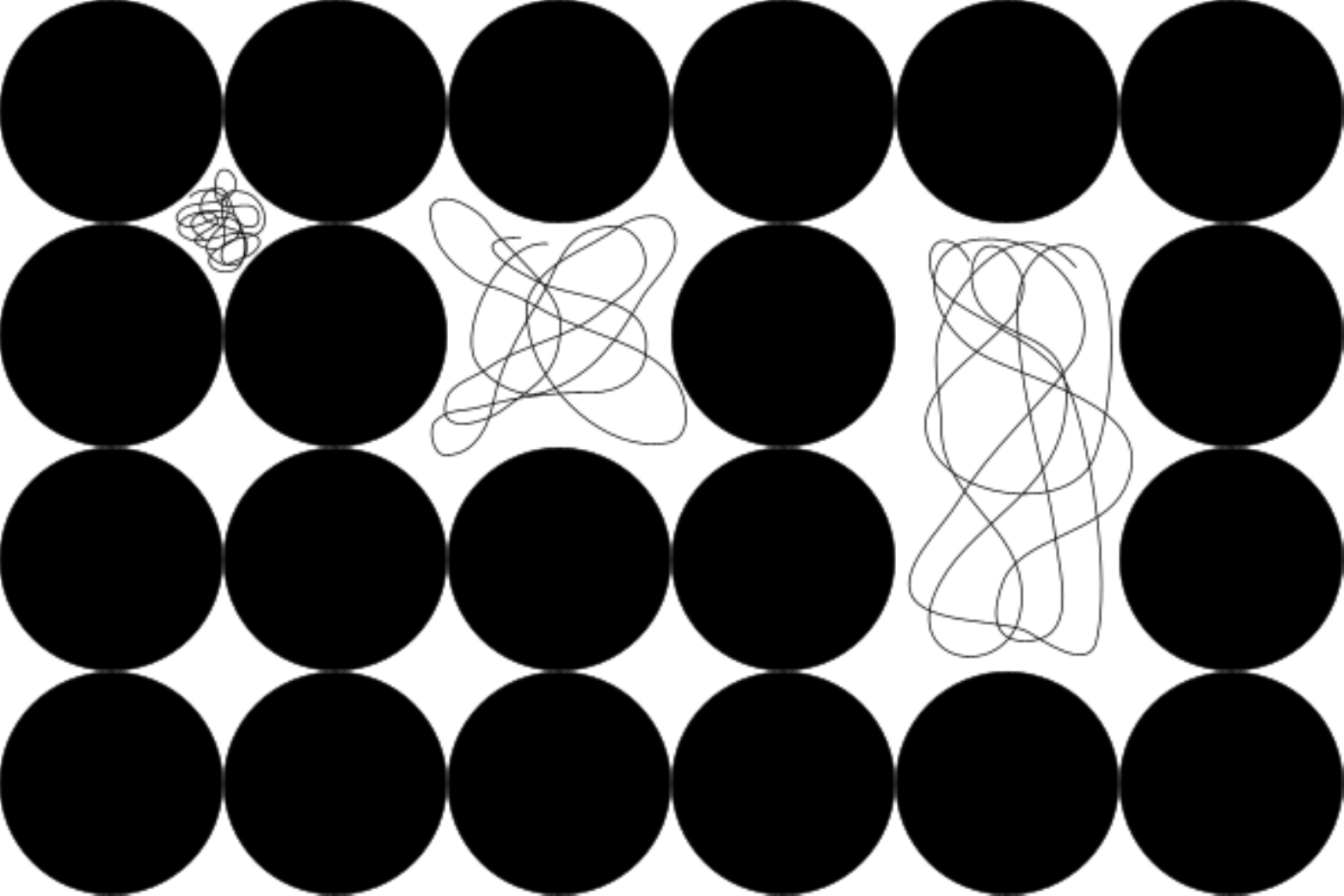}
    \end{minipage} & 
    \begin{minipage}{0.15\textwidth}
      \includegraphics[scale=0.15]{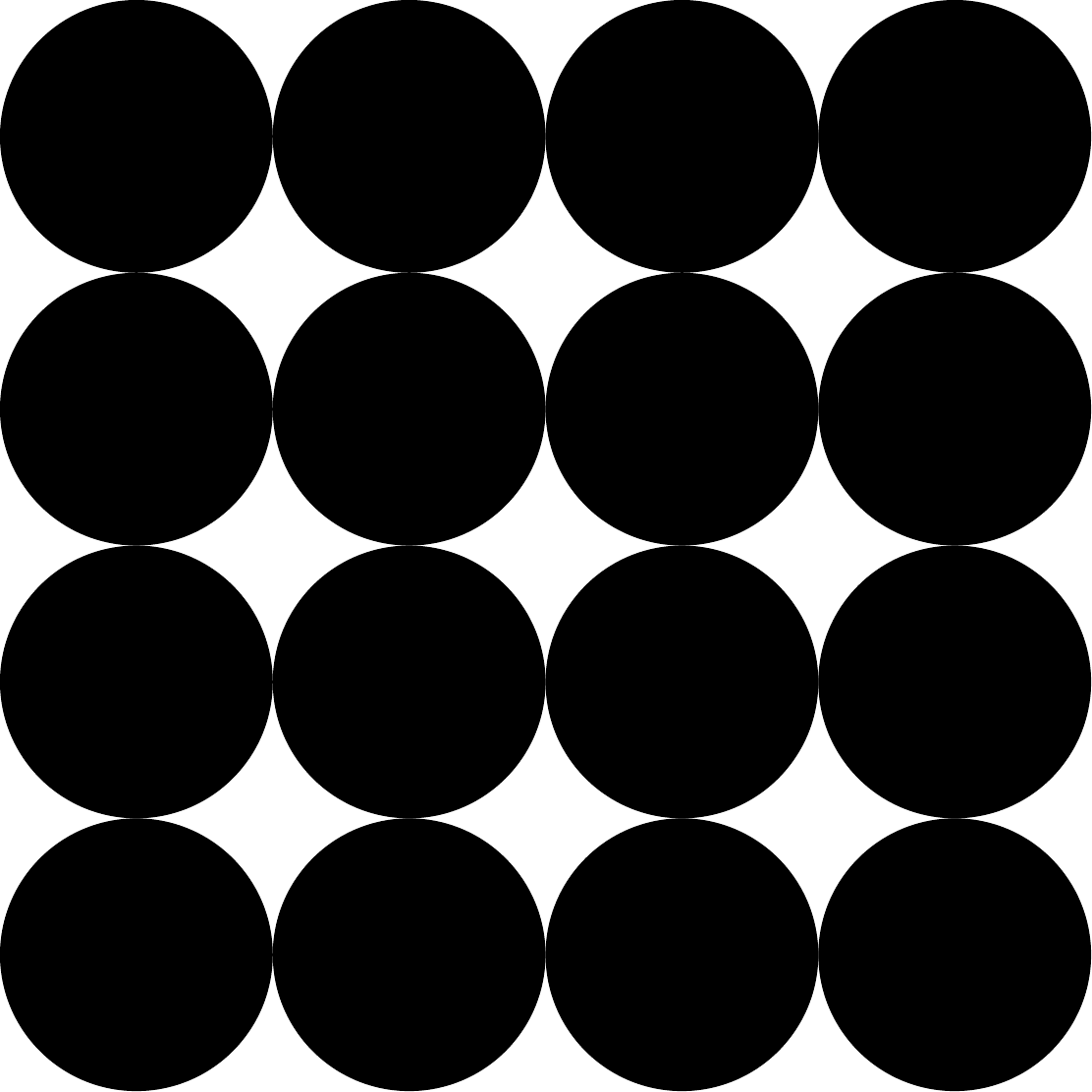}
    \end{minipage} \\
    (a) & (b) & (c) \\
  \end{tabular}
  \caption{Disorder realizations for increasing density of the background
  potential from left to right. (a) Low densities consist of single disks
  distributed in space. (b) In the intermediate- and high-density regimes there are
  holes of different sizes, whereas for a fully occupied lattice (c) there are only tiny holes left that cannot be occupied by disks of the potential.}
  \label{fig:densityIncreasePotential}
\end{figure}
The end-to-end distribution $P(r)$ for $\sigma_1$ and $\sigma_2$ shows this
behavior, which is the same for both algorithms. It is reflected by the double-peak
structure of $P(r)$ in Fig.~\ref{fig:singleDisorderConfigurationAnalysis}(b).
The small peak comes from the cavity where the polymer is pinned and the big one
from the nearby free space region, which is entropically much more favorable.
For the case of $\sigma_3$---no channel left between neighboring disks---the
distribution is characterized by a single peak, which corresponds to the
exploration of the tiny hole, containing pinpoint~1.  The broad single peaked
curve (solid black) in Fig.~\ref{fig:singleDisorderConfigurationAnalysis}(b) belongs
to pinpoint~2.  For all three cases of the diameter of the background
potential, the behavior is qualitatively the same. The large area around the pin
point is sampled by polymer configurations, leading to a broad end-to-end
distribution.  Figure~\ref{fig:singleDisorderConfigurationAnalysis}(c) shows the
tangent-tangent correlations for the different pinpoints. While pinpoint~2
leads to quick decorrelation of the tangents, which is characteristic for a free
polymer, things are completely different for pinpoint~1. $\sigma_1$ and
$\sigma_2$ show a correlation that is due to the fact that the polymer stretches
to the entropically favorable region next to pinpoint~1 through the channels
between the disks. This leads to a correlation on short to intermediate lengths
along the polymer. For $\sigma_3$, where no channels are left, the polymer
coils up in the cavity where it is pinned. This leads to strong
anti-correlations on short length scales. For both pinpoints, the two
employed simulation algorithms yield consistent results. This is reassuring
since neither continuum chain-growth algorithms nor our special multicanonical
method has been applied and tested before extensively.
\begin{figure*}\tt
  \begin{tabular}{ccc}
    \begin{minipage}{0.33\textwidth}
      \includegraphics[scale=1]{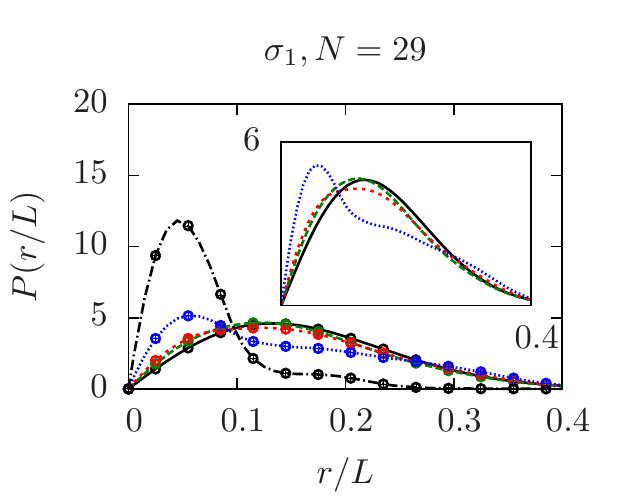}
    \end{minipage} &
    \begin{minipage}{0.33\textwidth}
      \includegraphics[scale=1]{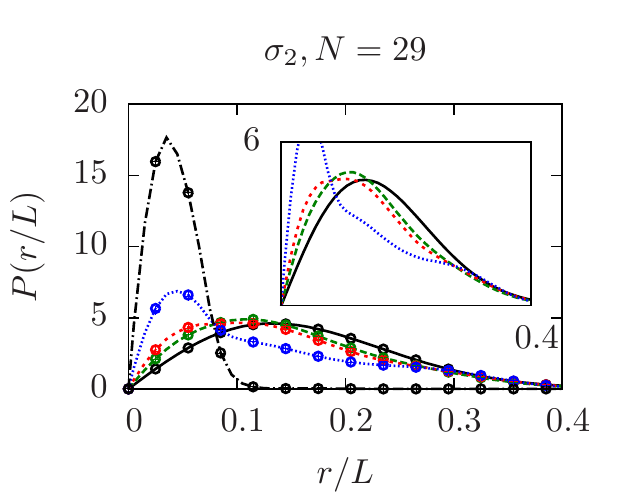}
    \end{minipage} &
    \begin{minipage}{0.33\textwidth}
      \includegraphics[scale=1]{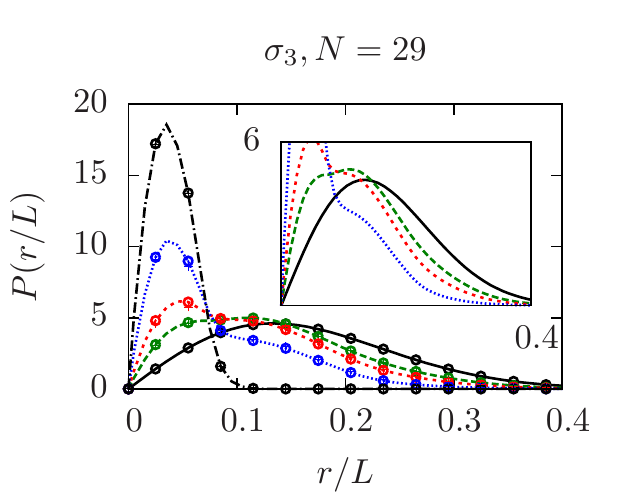}
    \end{minipage}\\
    (a) & (b) & (c)
  \end{tabular}
  \begin{minipage}{\textwidth}
    \caption{\label{fig:endToEndGrowthMucaHighDens}(Color online) End-to-end distribution
    function for site occupation $p=0.64$
    (\hspace{-0.2cm}\protect\includegraphics[scale=1]{greenLongDashed.pdf}, green), $0.76$
    (\hspace{-0.2cm}\protect\includegraphics[scale=1]{redDashed.pdf}, red), $0.89$
    (\hspace{-0.2cm}\protect\includegraphics[scale=1]{blueDotted.pdf}, blue),
    and
    $1$
    (\hspace{-0.2cm}\protect\includegraphics[scale=1]{blackDashDotted.pdf},
    black). The black solid curve is the end-to-end
    distribution of the free polymer as reference. The data marked by $\circ$ are
    from the growth algorithm and $+$ are from the multicanonical algorithm.
    The different plots are made for $\sigma_{1,2,3}$. The inset shows in each case
    the regime $p=0.64, 0.76, 0.89$ (the black solid curve is again the case
    $p=0$ as reference), where both the influence of the low-density regime and the influence of
    the small cavities play a role.} 
  \end{minipage}
\end{figure*}

After having looked at a single disorder realization that exhibits two exemplary cases, we
move on to the case of averaging over disorder. We take about 1500
disorder realizations for the quenched average.
We define the equality of the mean free path between neighboring disks and the mean
end-to-end distance of the free chain to mark the crossover between a low- and a
high-density regime. 
The effective free area per disk $A_{\mathrm{eff}}$ that is accessible
for the polymer is
\begin{equation}
  A_{\mathrm{eff}}=\frac{A-pM\sigma^2\pi/4}{pM},
  \label{eq:effFreeArea} 
\end{equation}
with $A$, $p$ and $\sigma$ as described in Sec.~\ref{sec:simulationParameters}
and $M$ the number of lattice sites---this is valid only for $\sigma \leq a$,
where neighboring disks do not overlap. The square root of $A_{\mathrm{eff}}$
gives the average free path per occupied site $x(p)$. The occupation $p_0$ where
$x(p)$ equals the  mean end-to-end distance of the polymer,
which is $\sqrt{\langle R_{\mathrm{ee}}^2 \rangle}=\sqrt{N}b$ for the free flexible case, marks the crossover
\begin{equation}
  p_0=(a/b)^2\frac{1}{1+\frac{\pi}{4}(\sigma/b)^2/N}\frac{1}{N}. 
\end{equation}
For the case considered here ($a=0.05,b=0.01,N=29$), this gives $p_0\approx
0.56$ for $\sigma=0.045$ and $p_0\approx 0.51$ for $\sigma=0.05$ and
$\sigma=0.051$.  
Figure~\ref{fig:endToEndGrowthMucaLowDens} shows the observables for a freely
jointed chain for low densities of the background potential. In this regime
($p\leq p_0$),
where the disorder landscape consists of free space and some randomly
distributed obstacles [see Fig.~\ref{fig:densityIncreasePotential}(a)], the
cases of different disk diameters $\sigma_{1,2,3}$ are similar. The
end-to-end distribution [Fig.~\ref{fig:endToEndGrowthMucaLowDens}(a)] is
characterized by a single peak that is shifted to the left and becomes more
pronounced for increasing density of the background, which can be interpreted as
compression of the polymer by the background potential. The tangent-tangent
correlations [Fig.~\ref{fig:endToEndGrowthMucaLowDens}(b)] show an
anti-correlation for increasing density of the background potential, which
goes quickly to zero correlation. This is characteristic for the
free polymer. 
\begin{figure*}\tt
  \begin{tabular}{ccc}
    \begin{minipage}{0.33\textwidth}
      \includegraphics[scale=1]{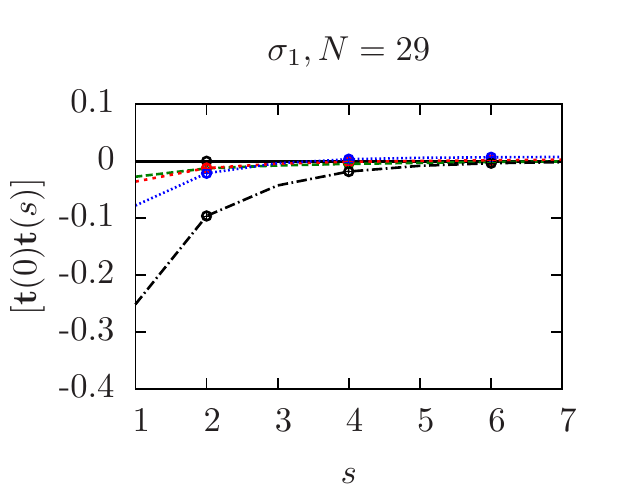}
    \end{minipage} &
    \begin{minipage}{0.33\textwidth}
      \includegraphics[scale=1]{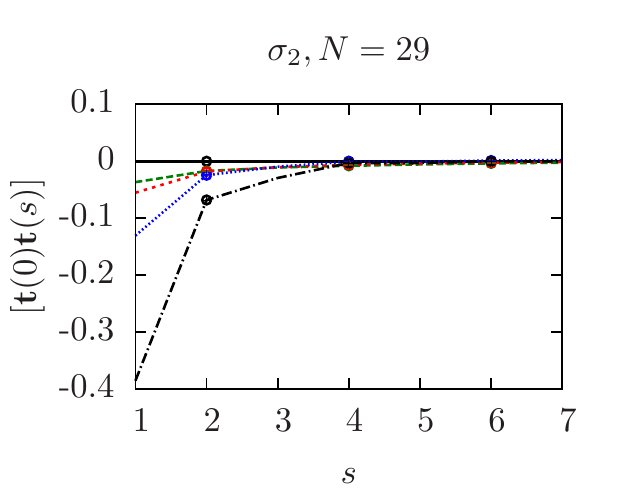}
    \end{minipage} &
    \begin{minipage}{0.33\textwidth}
      \includegraphics[scale=1]{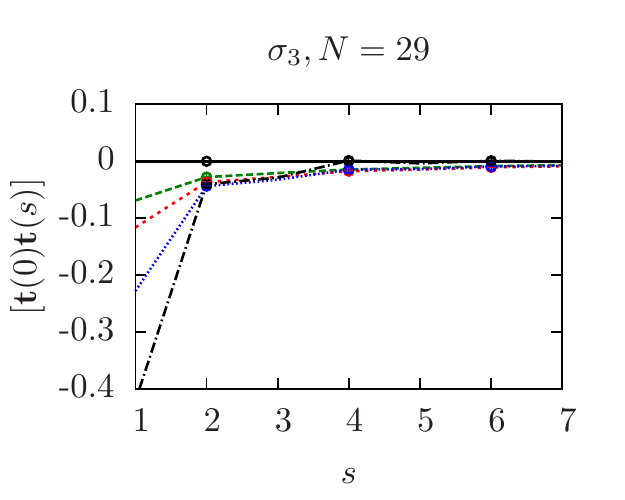}
    \end{minipage}\\
    (a) & (b) & (c)
  \end{tabular}
  \begin{minipage}{\textwidth}
    \caption{\label{fig:tanTanCorrAllDens}(Color online) Tangent-tangent
    correlations for $p=0.64$
    (\hspace{-0.2cm}\protect\includegraphics[scale=1]{greenLongDashed.pdf},
    green), $0.76$
    (\hspace{-0.2cm}\protect\includegraphics[scale=1]{redDashed.pdf}, red),
    $0.89$ (\hspace{-0.2cm}\protect\includegraphics[scale=1]{blueDotted.pdf},
    blue), and $1$
    (\hspace{-0.2cm}\protect\includegraphics[scale=1]{blackDashDotted.pdf},
    black). The black solid curve is for $p=0$ as reference. The data marked by
    $\circ$ are from the growth algorithm, whereas $+$ come from the
    multicanonical algorithm. (a), (b), and (c) differ in the disk diameter. The
    larger the disk diameter the stronger the anti-correlations on short length
    scales.} 
  \end{minipage}
\end{figure*}
The strength of the anti-correlation is one order of magnitude
weaker than for the case of high densities. The deviation of the mean square
end-to-end distance from the behavior of the
free case [dashed line in Fig.~\ref{fig:endToEndGrowthMucaLowDens}(c)] shows the
influence of the potential on the polymer in reducing the space to spread out.
The magnitude of the deviation from the free case is again insignificant compared to
the high-density case. Computationally we observe in the low-density regime
perfect agreement of the two simulation methods at the level of the
line-thickness in Fig.~\ref{fig:endToEndGrowthMucaLowDens}.

\begin{figure*}\tt
  \begin{tabular}{ccc}
    \begin{minipage}{0.33\textwidth}
      \includegraphics[scale=1]{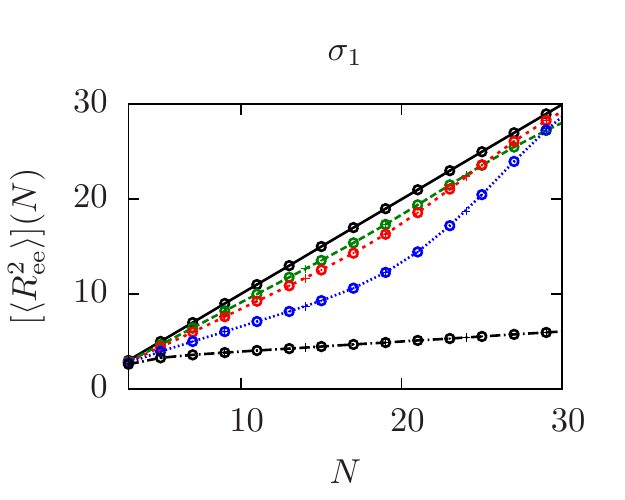}
    \end{minipage} &
    \begin{minipage}{0.33\textwidth}
      \includegraphics[scale=1]{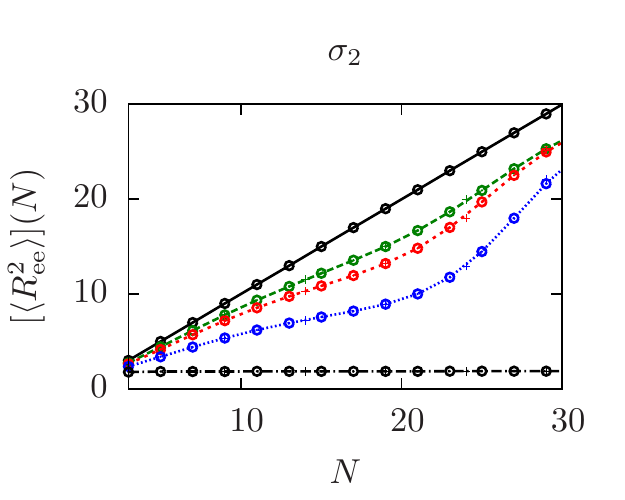}
    \end{minipage} &
    \begin{minipage}{0.33\textwidth}
      \includegraphics[scale=1]{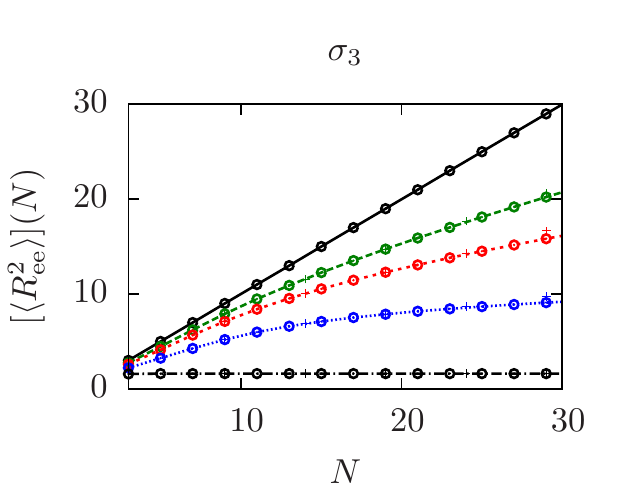}
    \end{minipage}\\
    (a) & (b) & (c)
  \end{tabular}
  \begin{minipage}{\textwidth}
    \caption{\label{fig:meanSquaredEndToEnd}(Color online) Mean square
    end-to-end distance (in units of squared bond length $b^2$) for $p=0.64$
    (\hspace{-0.2cm}\protect\includegraphics[scale=1]{greenLongDashed.pdf},
    green), $0.76$
    (\hspace{-0.2cm}\protect\includegraphics[scale=1]{redDashed.pdf}, red),
    $0.89$ (\hspace{-0.2cm}\protect\includegraphics[scale=1]{blueDotted.pdf},
    blue), and $1$
    (\hspace{-0.2cm}\protect\includegraphics[scale=1]{blackDashDotted.pdf},
    black) from top to bottom. The data marked by $\circ$ are from the growth
    algorithm; $+$ come from the multicanonical algorithm. The black solid curve
    shows the free polymer case which scales as $[\langle R_{\mathrm{ee}}^2
    \rangle] \propto N$.}  
  \end{minipage}
\end{figure*}

If we increase the density, the lattice structure dominates more and more,
which leads to a structure consisting of holes of different sizes
[Fig.~\ref{fig:densityIncreasePotential}(b)] that finally ends in a fully
occupied lattice where only tiny holes of space are left
[Fig.~\ref{fig:densityIncreasePotential}(c)]. 
The case of intermediate and high densities ($p > p_0$) is shown in
Figs.~\ref{fig:endToEndGrowthMucaHighDens},\ \ref{fig:tanTanCorrAllDens} and \ref{fig:meanSquaredEndToEnd}
for the three observables. The effect of cavities and channels dominates this
regime and leads to deviations depending on the choice of the diameter of
the disks of the background potential $\sigma_i$.

All three cases ($\sigma_{1,2,3}$) are determined by an interplay between
configurations where the pinpoint is inside a small cavity and configurations
whose pinpoint is in a larger area. For $p=1$ there finally are only small
cavities left. The case of $\sigma_3$, where the disks can overlap, is somewhat
special. For one thing, the occupation $p=0.59$ could play an
important role, as this is the site percolation threshold of the square lattice.
At this point, there is a percolating cluster in one direction which limits the
space for chain elongation. Furthermore, a polymer whose pinpoint is inside a cavity
cannot escape from it while this is possible for $\sigma_1$ and $\sigma_2$.
This effect can be well observed in the distribution of end-to-end distances.
For $p=0.64$ (long-dashed green curve in the plots of
Fig.~\ref{fig:endToEndGrowthMucaHighDens}), $\sigma_1$ and $\sigma_2$ still show
the low-density behavior, which is a single peak shifted to shorter lengths
compared to the free polymer. For $\sigma_3$, a small bulge next to the main
peak can be seen. The position of the bulge in the end-to-end distribution
corresponds to an extension of the chain of the order of $1$--$2$ bond lengths
which is the extent of the tiny holes
[Fig.~\ref{fig:densityIncreasePotential}(c)]. For intermediate densities it is
very probable that there is a larger free area next to a small cavity. A polymer
pinned inside a small cavity thus tries to escape from that region in order to
reach the entropically much more favorable space. Consequently, there is no
strong contribution from polymer configurations in small cavities.  This is of
course different for $\sigma_3$. 
For $p=0.76$ (short-dashed red curves), this effect enters also the case for $\sigma_1$
and $\sigma_2$ as there is less large space next to cavities. This reduces the
gain in entropy when leaving a cavity. This is more pronounced for $\sigma_2$ as
there the channels for escape are much smaller. For $p=1$, all
three cases yield qualitatively the same results again. In this case there is no more
benefit in escaping a small cavity, as there are only small cavities left.
For the cases of $\sigma_1$ and $\sigma_2$ the polymer thus stays in the
cavities whereas for $\sigma_3$ it cannot leave the cavity at all. 
The tangent-tangent correlations, Fig.~\ref{fig:tanTanCorrAllDens}, confirm the
findings for the end-to-end distribution. For high densities the polymer is
coiled up in a small region and therefore in a strongly folded state. This leads
to an anti-correlation of the tangents on very short length scales as in a
highly folded state it is more probable to have large angles between neighboring
bonds. However, this quickly averages out on longer length scales. This effect
gains importance with increasing density. A further effect, which is hardly seen
in the distributions of Fig.~\ref{fig:tanTanCorrAllDens} as the quenched
disorder average combines and thus smears different effects, is a stiffening of
the polymer---that is a positive correlation of bonds---on short length scales
for intermediate densities with $\sigma_1$ and $\sigma_2$. It can well be seen
for the single disorder configuration analysis in
Fig.~\ref{fig:singleDisorderConfigurationAnalysis}(c) (short-dashed red and
dotted blue curves) and
is already explained there.  A polymer that is pinned to a small hole that is
next to larger space stretches out to reach the entropically beneficial region
which leads to the above described positive correlations of tangents.  Cates and
Ball \cite{CatesBall1988} find similar effects due to energy instead of entropy
(tadpole configurations) with pinned polymers. 

The last thing to be discussed here is the mean square end-to-end distance in
dependence on the number of bonds $N$, which is shown in
Fig.~\ref{fig:meanSquaredEndToEnd}. The chain-growth algorithm produces
results in each step of growth which reduces the computational effort for
estimating the scaling of the mean square end-to-end distance. By comparison
with the multicanonical method we found that the
potential risk of systematic errors due to correlations of shorter and longer
chains can be neglected within our parameter range. For the multicanonical
method, the data for each polymer length have to be generated separately. This
leads to a higher computational effort in estimating the scaling of the
mean square end-to-end distance. For this reason we generated fewer data
points for the multicanonical method in Fig.~\ref{fig:meanSquaredEndToEnd}.

For the intermediate densities in Fig.~\ref{fig:meanSquaredEndToEnd}, both
algorithms again show the same behavior as described above. The surrounding
obstacles limit the extension of the polymer.  This effect increases for
increasing disorder density, which leads to a plateau in the mean square
end-to-end distance. This has also been found by Baumg\"artner and Muthukumar
\cite{Baumgaertner1987}. This effect dominates for the case of $\sigma_3$ where
neighboring disks leave no space for the polymer to escape
[Fig.~\ref{fig:meanSquaredEndToEnd}(c)]. Things are different for the cases of
$\sigma_1$ and $\sigma_2$ [Figs.~\ref{fig:meanSquaredEndToEnd}(a) and
\ref{fig:meanSquaredEndToEnd}(b)].  While
the first part of the curves shows the same behavior, an increase of the mean
square end-to-end distance for increasing number of bonds with a slope $m < 1$,
this suddenly changes to a steep slope with $m > 1$.  The slope $m$ larger than
1 is due to the reduced angular interval that is available after the polymer has
left a small cavity through a narrow channel.  Accordingly the polymer is forced
in a certain direction, which increases its extension. 

\begin{figure}\tt
  \includegraphics[scale=0.35]{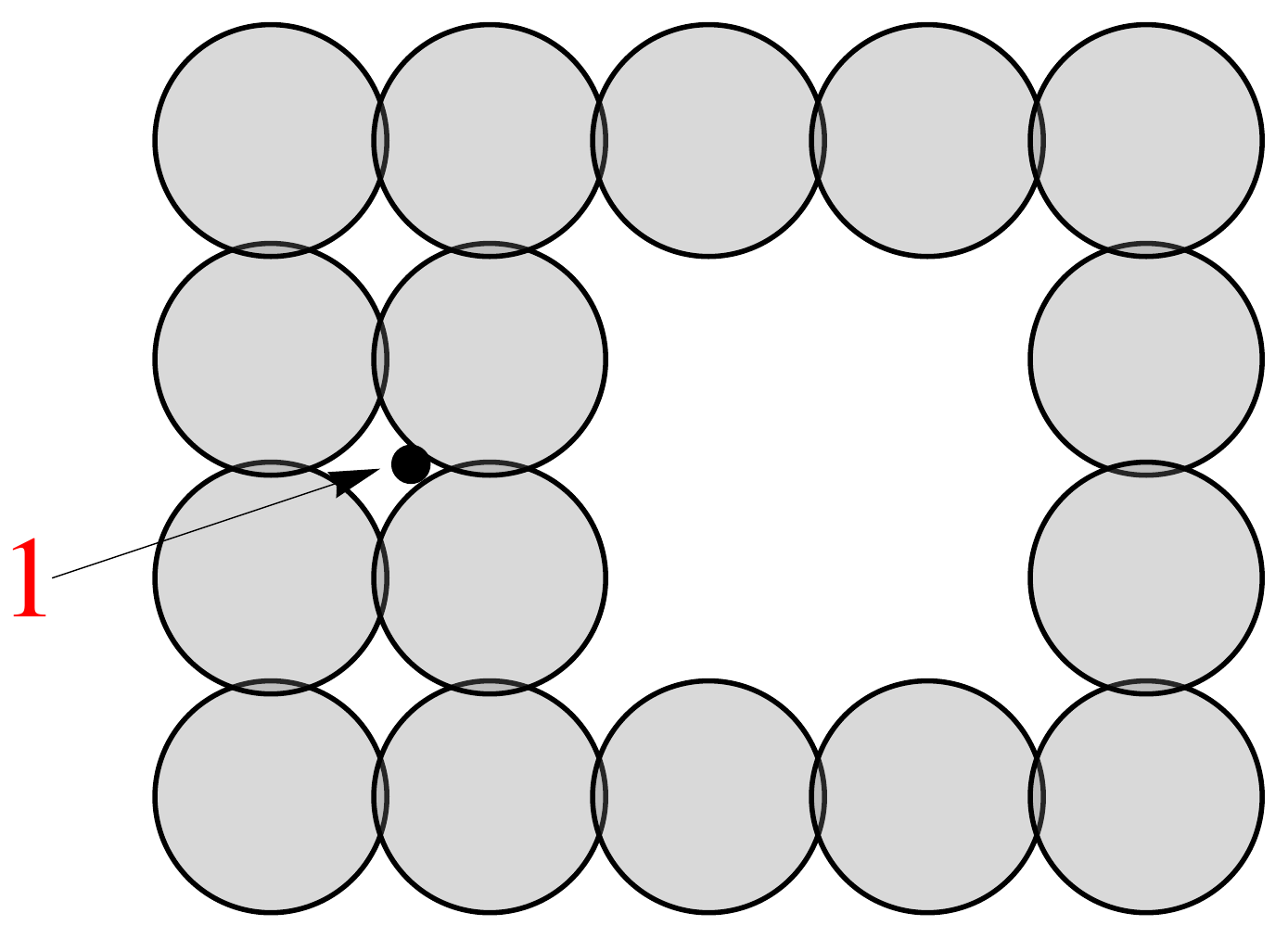} 
  \caption{\label{fig:mucaProbCav} Hard-disk configuration and pinpoint~1 for
  $\sigma_3$. For those cases the multicanonical method hardly converges as it
  mainly explores the area outside the cavity.} 
\end{figure}
After having described the phenomenology of the problem, we want to
comment briefly on the algorithms for the above described problem. For the cases of
$\sigma_1$ and $\sigma_2$ the two algorithms produce fully
consistent results.  While for shorter chain lengths the two algorithms also
agree for $\sigma_3$ for high density, they start showing small deviations from
each other for $N=29$ which is barely visible only in
Fig.~\ref{fig:meanSquaredEndToEnd}(c).  In analyzing the deviations, we found that
this effect increases for increasing chain length. The deviations occur if a pinpoint is
in a corner of a small cavity next to a larger free area.  The growth algorithm
explores the nearby region, building up a dense network of polymers by growing
them in parallel. The multicanonical routine, however, explores space by
updating an existing configuration, thereby taking into account overlaps with
the surrounding disks. Afterward, these overlapping configurations are
calculated out by the reweighting process. In a case as depicted in
Fig.~\ref{fig:mucaProbCav} there are some difficulties with this process. The
space for configurations that are allowed is relatively small.  As the
multicanonical routine is not restricted to the allowed region, the sampling in
the allowed region is very rare, which leads to convergence problems for the
case of small cavities. 

\section{Conclusion and outlook}\label{sec:conc}

We investigated the phenomenology of a flexible polymer exposed to a quenched
disorder landscape consisting of hard disks placed on the sites of a square
lattice. We recovered the result found by Baumg\"artner and Muthukumar
\cite{Baumgaertner1987} that the polymer shrinks in the presence of obstacles,
which is reflected in a high probability for chains with a short end-to-end
distance in the end-to-end distribution function. For high densities we found a
dependency of the characteristics of the polymer on the microscopic structure of
the disorder. Obstacles that leave no channels between neighboring sites lead to
a plateau for all densities in the scaling of the mean square end-to-end
distance for increasing numbers of bonds. This effect is inverted for the case
of channels between neighboring disks. The mean square end-to-end distance shows
a steep slope for intermediate to high densities as long as the polymer benefits
from the channels by being able to explore entropically favorable regions.

We cross-checked our findings by applying two conceptually very different
simulation algorithms: a continuum chain-growth algorithm with population
control and the multicanonical method based on Markov chains in a generalized
ensemble. In doing so we found very good agreement in almost all situations.
Only for certain cases of high densities and subtle structures did we encounter
problems with the multicanonical method.

After having checked our methods within a system that could be well controlled,
we mean to apply them to more sophisticated disorder landscapes such as hard-disk
fluids. Also, the polymer model can be adapted such that it includes bending
terms which makes it applicable to biological polymer systems. Those are often 
modeled by the worm-like chain. 

\section*{Acknowledgments}

We gratefully thank Sebastian Sturm, Niklas Fricke, and Mathias Aust for
discussion and beneficial advice about the problem addressed here. The project was
financially supported by the Leipzig Graduate School of Excellence GSC 185
``BuildMoNa'' and the Deutsch-Franz\"osische Hochschule (DFH-UFA). 

\bibliographystyle{apsrev4-1.bst}

\end{document}